\documentclass[showpacs,amsmath,amssymb,12pt]{revtex4}

\usepackage{graphicx}
\usepackage{dcolumn}
\usepackage{bm}
\usepackage{subfigure}
\makeatother

\begin{document}

\title{About Starobinsky Inflation}

\author{Daniel M\"uller}
\email{muller@fis.unb.br}
\affiliation{Instituto de F\'\i sica, UnB \\ 
Campus Universit\'ario Darcy Ribeiro\\ 
Cxp 04455, 70919-970, Brasília DF \\
Brasil}
\author{Sandro D. P. Vitenti}
\email{vitenti@isoftware.com.br}
\affiliation{Instituto de F\'\i sica, UnB \\ 
Campus Universit\'ario Darcy Ribeiro\\ 
Cxp 04455, 70919-970, Brasília DF \\
Brasil}
\date{\today}

\begin{abstract}        
It is believed that soon after the Planck era, space time should have a
semi-classical nature. According to this, the escape from General Relativity theory is 
unavoidable. Two geometric counter-terms are needed to regularize the divergences which come from 
the expected value. These counter-terms are responsible for a higher
derivative metric gravitation. Starobinsky idea was that
these higher derivatives could mimic a cosmological constant. In this
work it is considered numerical solutions for general Bianchi I anisotropic space-times in
this higher derivative theory. The approach is ``experimental'' in the
sense that there is no attempt to an analytical investigation of the
results. It is shown that for zero cosmological
constant $\Lambda=0$, there are sets of initial conditions which form 
basins of attraction that asymptote
Minkowski space. The complement of this set of initial conditions form
basins which are attracted to some singular solutions. It is also
shown, for a cosmological constant $\Lambda> 0$ that there are
basins of attraction to a specific de Sitter solution. This result 
is consistent with Starobinsky's initial idea. The complement of this
set also forms basins that are attracted to some type of singular
solution. 
Because the singularity is characterized by curvature scalars, it must
be stressed that the basin structure obtained is a topological
invariant, i.e., coordinate independent.
\end{abstract}
\pacs{98.80.Cq, 98.80.Jk, 05.45.-a}
                            
\maketitle

\section{Introduction} 
The semi-classical theory consider the back reaction of quantum fields in a classical geometric 
background. It began about forty years ago with De Witt \cite{DeWitt}, and since then, its 
consequences and applications are still under research, see for example \cite{Hu}.

Different of the usual Einstein-Hilbert action, the predicted  gravitational action allows differential 
equations with fourth order derivatives, which is called the full theory \cite{DeWitt}, see also 
\cite{liv}. 

There are several problems connected to the full theory which prove that it is not consistent with our 
expectation of the present day physical world, see \cite{Simon} and
references therein. In this work, Simon and Parker-Simon suggest a
perturbative approach which is very interesting.

The full higher order theory was previously studied by Starobinsky \cite{S}, and more recently, also 
by Shapiro, Pelinson and others \cite{Shapiro}. Starobinsky idea was that the higher order terms 
could mimic a cosmological constant. In \cite{Shapiro} only the homogeneous and isotropic space 
time is studied.

The full theory
with four time derivatives is addressed, which apparently was first
investigated in Tomita's article \cite{berkin} for general Bianchi I
spaces. They found that the presence of anisotropy contributes to the 
formation of the singularity. Berkin's work shows that a quadratic
Weyl theory is less stable than a quadratic Riemann scalar $R^2$.
Barrow and Hervik found exact and analytic solutions for anisotropic
quadratic gravity with $\Lambda > 0$ that do not approach a de Sitter
space time. In that article Barrow and Hervik discuss Bianchi
types $II$ and $VI_h$ and also a very interesting stability criterion concerning 
small anisotropies. There is also a recent article by Clifton and Barrow in 
which Kasner type solutions are addressed. H. J. Schmidt does a recent and very interesting  
review of higher order gravity theories in connection to cosmology \cite{hjs}.

On the other hand, numerical investigations of initial conditions, was
already addressed in Einstein's theory of gravitation for the homogeneous
Bianchi IX solutions. Basin of attraction to asymptotic solutions, and
the fractal nature of the basin boundary are strong indications that
the early stages of the Universe were very chaotic, according to
Einstein's theory, see for example \cite{cornish}.

For general anisotropic Bianchi I homogeneous space times, the full
theory reduces to a system of nonlinear ordinary differential
equations. The numerical solutions for this system with derivatives of fourth order in time
was previously obtained by us \cite{sandro}. 

In this present work a numerical analysis of initial conditions for
this nonlinear system mentioned above, is given.

Only the vacuum energy momentum classical source is considered in the full theory. It should be valid 
soon after the Planck era in which vacuum classical source seems the most natural condition.
The chaotic nature of the singularity is not addressed. 

There is not a unique way to totally characterize a singularity. In
this numerical investigation, the singularity is defined when 
$R_{ab}R^{ab}>10^6$.

It is found that for zero cosmological constant $\Lambda=0$, there are
basins that are   
attracted to zero, constant $4$-curvature solutions. This proves that
according to this theory, 
there are regions in the space of initial conditions, for which the
Minkowski geometry is structurally stable. That is, for considerably
anisotropic initial conditions isotropisation occurs,
and Minkowski space is obtained asymptotically, without the need for
fine tuning. 

On the other hand for a nonzero cosmological constant $\Lambda > 0$
the basins attract to a specific de Sitter geometry, which is a constant
homogeneous, non null $4$-curvature solution. Thus indicating that 
Starobinsky's inflation is structurally stable in the above sense.

Also, the complement of the set of initial conditions for the stable,
physically accepted solutions, for both cases $\Lambda=0$ and $\Lambda
> 0$, form basins of attraction to 
singular, in some sense, Universes. Thus physically acceptable initial
conditions evolve in very few Planck times, to Universes with very high curvature
scalars. In this sense, this theory certainly is not a complete
one. 

It should be mentioned that the isotropisation process for non zero
cosmological constant is not a peculiarity of these higher derivatives
theories. It occurs in ordinary Einstein General Relativity theory. 
This is proved in the very interesting article \cite{wald}, where 
for $\Lambda >0$ all Bianchi models, except for highly positively curved
Bianchi IX become asymptotically de Sitter. The remarkable difference
between General Relativity theory and this higher derivatives theories
is that the isotropisation depends on the initial conditions. Isotropisation
is strongly dependent on initial conditions for quadratic curvature
theories such as \eqref{acao}.

The following conventions and unit choice are taken $R^a_{bcd}=\Gamma^a_{bd,c}-...$, 
$R_{ab}=R^c_{acb}$, $R=R^a_a$, metric signature $-+++$, Latin symbols run from $0-3$, Greek 
symbols run from $1-3$ and $G=\hbar=c=1$.  

\section{The Model}

The Lagrange function is, 
\begin{equation}
{\cal L}=\sqrt{-g}\left[\Lambda + R+
\alpha\left( R^{ab}R_{ab}-\frac{1}{3}R^2\right)+\beta R^2\right] {\label{acao}} + {\cal L}_c\,.
\end{equation}
Metric variations in the above action results in
\begin{equation}
E_{ab}=G_{ab}+\left(\beta-\frac{1}{3}\alpha\right)H^{(1)}_{ab}+\alpha H^{(2)}_{ab}
-T_{ab}-\frac{1}{2}\Lambda g_{ab},
\label{eqtotal}
\end{equation}
where
\begin{eqnarray}
&&H^{(1)}_{ab}=\frac{1}{2}g_{ab}R^2-2RR_{ab}+2R_{;ab}-2\square Rg_{ab},\\
&&H^{(2)}_{ab}=\frac{1}{2}g_{ab}R_{mn}R^{mn}+R_{;ab}-2R^{cn}R_{cbna}-\square R_{ab}-\frac{1}{2}\square Rg_{ab}, \\
&&G_{ab}=R_{ab}-\frac{1}{2}g_{ab}R,
\end{eqnarray}
and $T_{ab}$ is the energy momentum source, which comes from the classical part of the Lagrangian 
  ${\cal L}_c$. Only vacuum solutions $T_{ab}={\cal L}_c=0$  will be considered in this paper since it seems 
the most natural condition soon after the Planck era.
 
The covariant divergence of the above tensors are identically zero due their variational definition. The 
following Bianchi Type I line element is considered
\begin{equation}
ds^2=-dt^2+[a_1(t)]^2dx^2+[a_2(t)]^2dy^2+[a_3(t)]^2dz^2 \label{elinha},
\end{equation}
which is a general spatially flat and anisotropic space, with proper time $t$. With this line element all 
the tensors which enter the expressions are  diagonal. The substitution of \eqref{elinha} in 
\eqref{eqtotal} with $T_{ab}=0$, results for the spatial part of \eqref{eqtotal}, in differential 
equations of the type 
\begin{eqnarray}
&&\frac{d^{4}}{dt^4}a_1=f_1\left( \frac{d^3}{dt^3}a_i,\ddot{a}_i,\dot{a}_i,a_i
\right) \label{edo1}\\
&&\frac{d^{4}}{dt^4}a_2=f_2\left( \frac{d^3}{dt^3}a_i,\ddot{a}_i,\dot{a}_i,a_i
\right)\\
&&\frac{d^{4}}{dt^4}a_3=f_3\left( \frac{d^3}{dt^3}a_i,\ddot{a}_i,\dot{a}_i,a_i
\right)\label{edo3},
\end{eqnarray}
where the functions $f_i$ involve the $a_1,\;a_2,\;a_3$, and their
derivatives in a polynomial fashion, see the Appendix \ref{ap1}. The 
very interesting article \cite{Noakes} shows that the theory which follows from \eqref{acao} has a 
well posed initial value problem. This question has some similarities
which General Relativity theory since it is necessary to solve the
problem of boundary conditions and initial conditions
simultaneously. According to Noakes \cite{Noakes}, once a consistent boundary condition is chosen 
initially, the time evolution of the system is uniquely specified. 
In \cite{Noakes} the differential
equations for the metric are written in a form suitable for the
application of the theorem of Leray \cite{Leray}. The numerical
solutions of the partial differential equations in \eqref{eqtotal} are
not going to be addressed in this present work.

For homogeneous spaces the differential equations in
\eqref{eqtotal} reduce to non linear ordinary differential equations.
Then, instead of going through the general construction given in
\cite{Noakes}, in this particular case, the 
existence and uniqueness of the solutions of \eqref{edo1}-\eqref{edo3}
reduce to the well known problem of existence and uniqueness of
solutions of ordinary differential equations. For a proof on local
existence and uniqueness of solutions of differential equations see, for instance \cite{reedsimon}.

Besides the equations \eqref{edo1}-\eqref{edo3}, we have the temporal part of \eqref{eqtotal}. To 
understand the role of this equation we have first to study the covariant divergence of the equation 
\eqref{eqtotal}, $$\nabla_aE^{ab} = \partial_aE^{ab} + \Gamma^a_{ac}E^{cb} + 
\Gamma^{b}_{ac} E^{ac} = 0.$$  Remind that the coordinates being used
are \eqref{elinha} and that $T_{ab}=0$. Since the differential
equations \eqref{edo1}-\eqref{edo3} are solved numerically $E^{\alpha \alpha}\equiv 0,$ 
\begin{eqnarray}
&&\partial_0E^{00} + \Gamma^a_{a0} E^{00} + 
\Gamma^{0}_{00} E^{00} = 0, \nonumber\\
&&E^{00}(t)\equiv 0, 
\label{vinculo}
\end{eqnarray}
where $E_{00}$ is the $00$ component of \eqref{eqtotal}.
If $E_{00}=0$ initially, it will remain zero at any instant. 
Therefore the equation $E_{00}$ acts as a constraint on the initial conditions and we use it to test the 
accuracy of our results.

For a space-like vector $v^\alpha$ and a time-like vector $t^a=(1,0,0,0)$ tidal forces are given by 
the geodesic deviation equations
\begin{eqnarray*}
&&t^a\nabla_a v^\alpha=R^\alpha_{mn\beta}t^mt^nv^\beta \\
&&t^a\nabla_a v^\alpha=R^\alpha_{00\beta}v^{\beta}\\
&&R^\alpha_{00\beta}=\delta_{\alpha\beta}\left(\frac{\ddot{a}_\alpha}{a_\alpha}\right).
\end{eqnarray*}
The theory predicted by \eqref{acao} is believed to be correct if the tidal forces are less than $1$ in 
Plank units units, 
\begin{equation}
|R^\alpha_{0\alpha0}|\leq 1\;\; \text{(no summation.)}\label{mare}
\end{equation}
When this condition is not satisfied, quantum effects could introduce further modifications into 
\eqref{acao}.

\section{Analysis of initial conditions} 
Ordinary differential equations are deterministic in the sense that
initial conditions specify the dynamical evolution of the system
uniquely. 

The interest in the numerical investigation is in the sensibility upon
initial conditions. In this work, the system is treated as an exit
system. 

It is a relatively simple approach in which asymptotic classes of
solutions are recognized. The initial conditions are identified
with respect to which asymptotic solution the system evolves to \cite{ott}. For
this purpose, holes are cut in the phase space and if the system
falls into a specific hole, this specifies the asymptotic class. The
closure of the set of initial conditions which asymptotes a given solution
is called the basin of attraction.   

The presence of basins of attraction is an indication of structural
stability, since for many initial conditions, without the need for
fine tuning, the same class of asymptotic solution is obtained. It is
also a topological invariant, since a smooth coordinate transformation is impossible 
to modify the Hausdorff dimension of a set \cite{ott}.

In subsections \ref{A} and \ref{B}, the differential equations
\eqref{edo1}-\eqref{edo3} are numerically solved and the initial conditions are chosen in the following way. All the higher 
derivatives of the metric set initially to zero 
\begin{equation}
\frac{d^3}{dt^3}a_\alpha(t=0)=\ddot{a}_\alpha(t=0)=0\;\; \text{and} \;
a_1(t=0)=a_2(t=0)=a_3(t=0)=1 \label{ci1}
\end{equation}
The { \it Hubble  constants} in each direction, Appendix \ref{ap1}, 
\begin{equation}
H_1=\frac{\dot{a}_1}{a_1}\neq 0,\;\; H_2=\frac{\dot{a}_2}{a_2}\neq,
0 \label{ci2}
\end{equation}
together with the Hamiltonian constrain \eqref{vinculo} fix the initial value of 
\begin{equation}
H_3=\frac{\dot{a}_3}{a_3}. \label{ci3}
\end{equation}
This value is not unique since eq. \eqref{vinculo} in the coordinates \eqref{elinha} form a
polynomial of degree $3$ in the first derivatives. For arbitrary initial values for $H_1$ and
$H_2$ there could be at most $3$ real distinct initial values for
$H_3$. One of these values is chosen. It can be seen that the initial
values are consistent with the condition \eqref{mare}, resulting in
null tidal forces.

\subsection{Stability of Minkowski space \label{A}}
\begin{figure}
\includegraphics[scale=0.8]{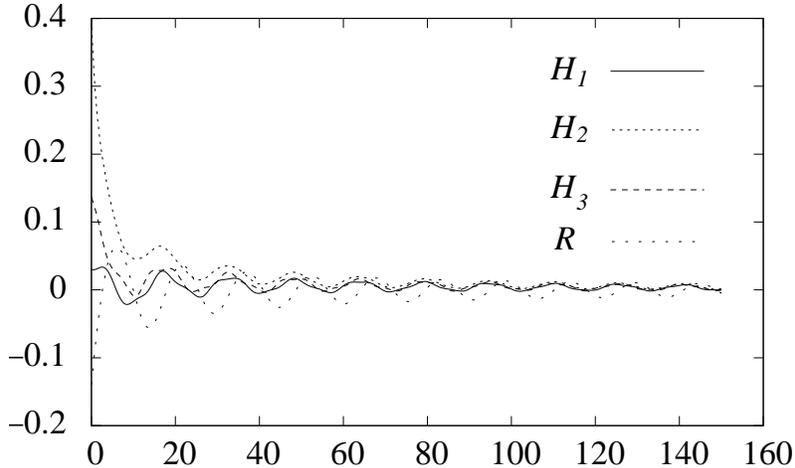}
\caption{Numerical solution of \eqref{edo1}-\eqref{edo3}, with the initial condition described in
eqs. \eqref{ci1}-\eqref{ci3}, with $\alpha=1$, $\beta=-1$ and $\Lambda=0$, plotted 
against time, in Planck units. It can be seen the isotropisation towards Minkowski
space. \label{mink}}
\end{figure}

\begin{figure}
\includegraphics[scale=0.8]{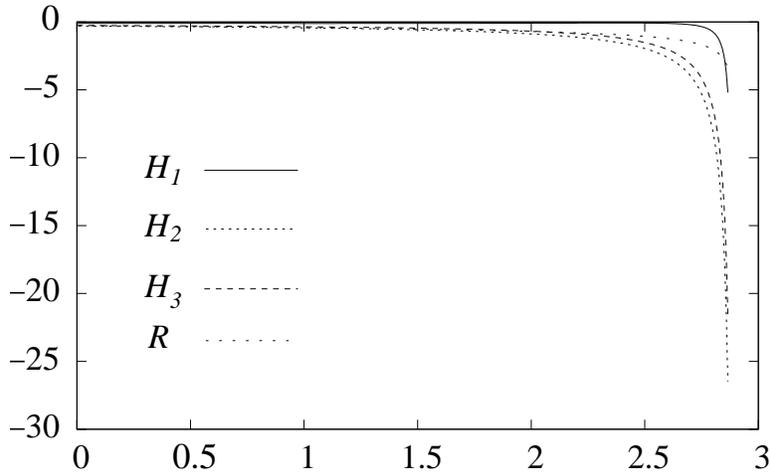}
\caption{Numerical solution of \eqref{edo1}-\eqref{edo3}, with the initial condition described in
eqs. \eqref{ci1}-\eqref{ci3}, with $\alpha=1$, $\beta=-1$ and $\Lambda=0$, plotted 
against time, in Planck units. It can be seen the formation of a
singularity with $H_1<0$ and  $H_2<0$ and $H_3<0$. \label{singw}}
\end{figure}

\begin{figure}
\includegraphics[scale=0.8]{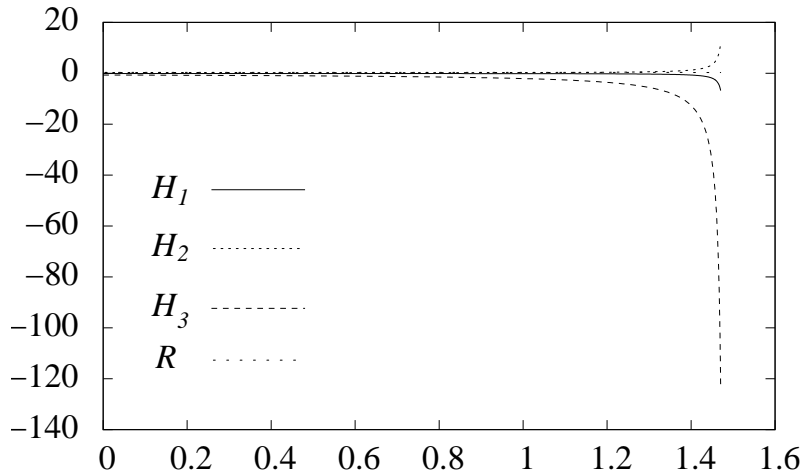}
\caption{Numerical solution of \eqref{edo1}-\eqref{edo3}, with the initial condition described in
eqs. \eqref{ci1}-\eqref{ci3}, with $\alpha=1$, $\beta=-1$ and $\Lambda=0$, plotted 
against time, in Planck units. It can be seen the formation of a
singularity with at least $H_1>0$ or $H_2>0$ or $H_3>0$, and at least one of the $H$ is 
negative. \label{singgr}}
\end{figure}

\begin{figure}
\includegraphics[scale=0.4]{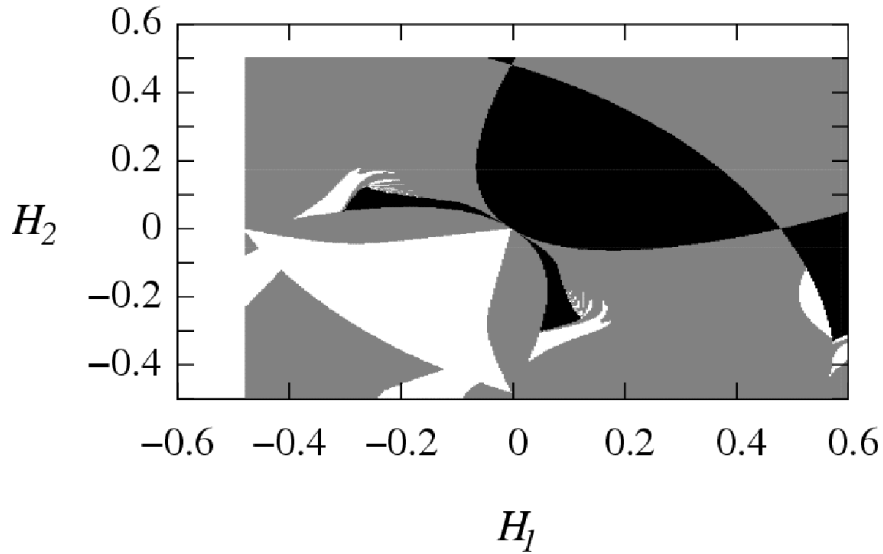}
\caption{A two dimensional section defined by \eqref{ci1}-\eqref{ci3} and 
\eqref{vinculo}, for the phase space of \eqref{edo1}-\eqref{edo3},
with $400\times 400$ initial values for $H_1$ and $H_2$. $\alpha=1$, $\beta=-1$ and $\Lambda=0$. 
The black regions are the basins of attraction to the Minkowski
space. The white and gray regions are initial conditions which
asymptote the singularities shown in FIG. \ref{singw} and
FIG. \ref{singgr} respectively.\label{stm}}
\end{figure}

In this subsection the cosmological constant is set to zero $\Lambda=0$. 
The numerical solutions of \eqref{edo1}-\eqref{edo3} are identified according to their asymptotic classes.
In FIG. \ref{mink} it is shown the isotropisation process toward
Minkowski space. 
A singularity for which all the 3 dimensions shrink to zero, with 
$H_1<0$, $H_2<0$ and $H_3<0$ when $t\rightarrow \infty$ is shown in
FIG. \ref{singw}.
In FIG. \ref{singgr} it is shown a singularity for which one of the dimensions increases, either $H_1>0$
or $H_2>0$ or $H_3>0$ and at least one of the $H$ is negative, when $t\rightarrow \infty$.
In FIG. \ref{stm},  a {\it black},  {\it white} or {\it gray} point is addressed to a
initial condition whether it falls into the classes shown in
FIG. \ref{mink}, FIG. \ref{singw} or FIG. \ref{singgr}
respectively.
\subsection{Stability of de Sitter space \label{B}}
\begin{figure}
\includegraphics[scale=0.4]{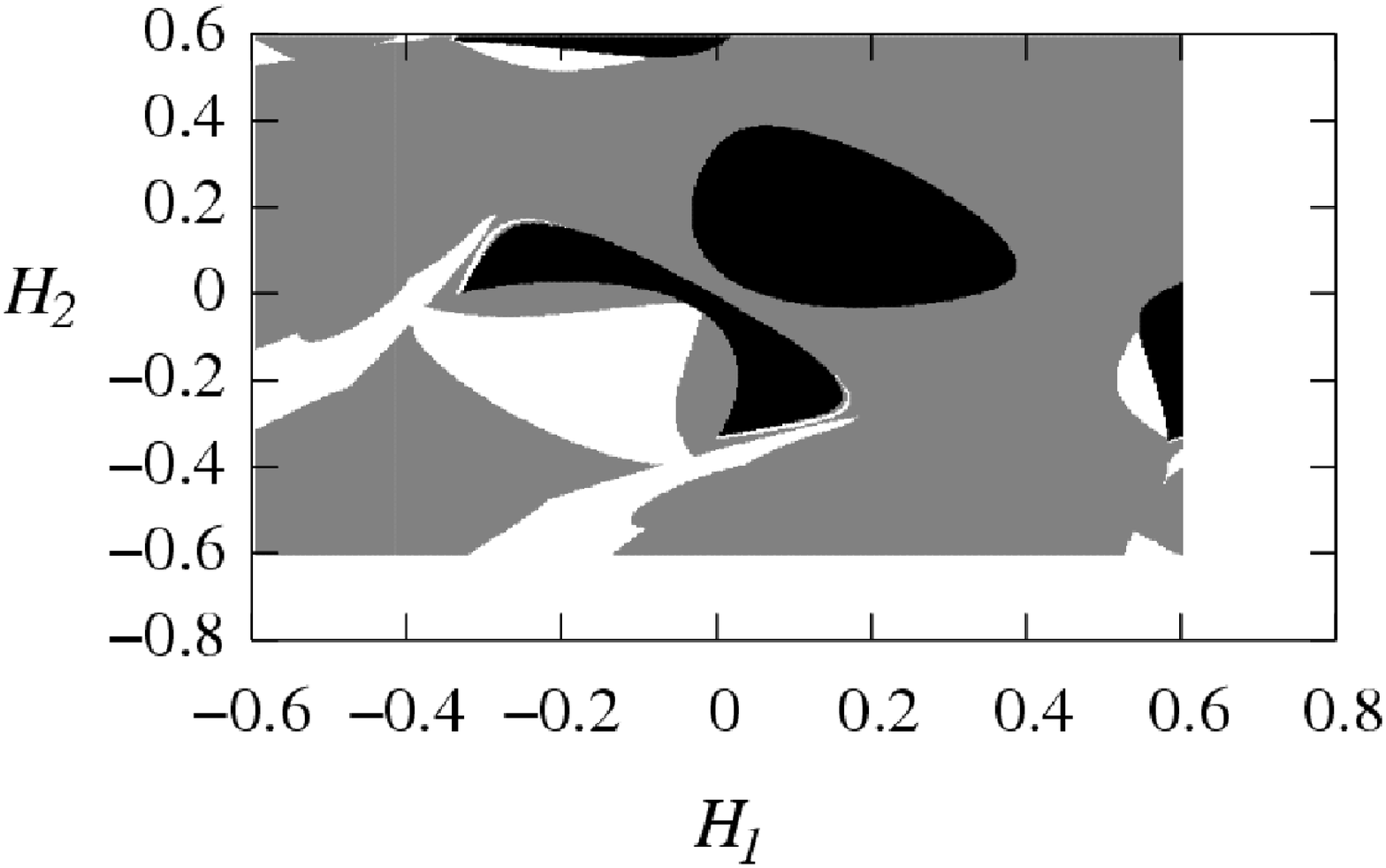}
\caption{A two dimensional section defined by \eqref{ci1}-\eqref{ci3} and 
\eqref{vinculo}, for the phase space of \eqref{edo1}-\eqref{edo3},
with $400\times 400$ initial values for $H_1$ and $H_2$. 
$\alpha=1$, $\beta=-1$ and $\Lambda=0.02$. The black regions are the
basins of attraction to the de Sitter space sown in
FIG. \ref{isotrop}. The white and gray regions are initial conditions which
asymptote the singularities shown in FIG. \ref{singw} and FIG. \ref{singgr} respectively.\label{std}}
\end{figure}

\begin{figure}
\includegraphics[scale=0.8]{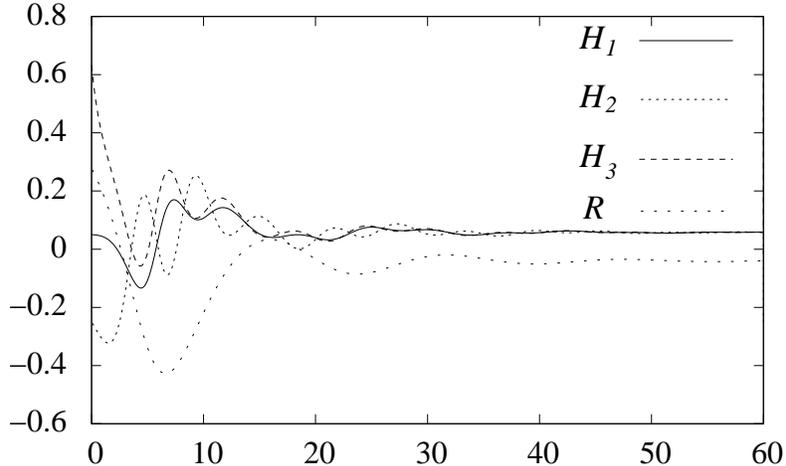}
\caption{Numerical solution of \eqref{edo1}-\eqref{edo3}, with the initial condition described in
eqs. \eqref{ci1}-\eqref{ci3}, with $\alpha=1$, $\beta=-1$ and $\Lambda=0.02$, plotted 
against time, in Planck units. It can be seen the isotropisation
toward a de Sitter space with $R=R_{ab}g^{ab}=-0.04$. \label{isotrop}}
\end{figure}

In this subsection a non zero cosmological constant is chosen
$\Lambda=0.02$. The map of initial conditions for the differential
equations \eqref{edo1}-\eqref{edo3}, is almost identical to
the last section, except for the {\it black} point. Each black point
in FIG. \ref{std} corresponds to initial conditions whose solution asymptote a constant $4$
curvature space $R=R_{ab}g^{ab}=-0.04$ shown in FIG. \ref{isotrop}.
In FIG. \ref{isotrop} the time evolution of a {\it black point} in FIG. \ref{std} shows
the isotropisation process toward a de Sitter, constant curvature solution.
The basins of attraction to this same de Sitter solution, is an
indication of stability in some sense.
\newpage 
\section{Conclusions}
In the present work it is considered general anisotropic Bianchi I
homogeneous space times given by the line element \eqref{elinha}. For
this line element, the theory given in \eqref{acao}, reduces to a
system of ordinary nonlinear differential equations with four time
derivatives. This system is 
numerically investigated. In particular, an analysis of initial
conditions is given, which is the interest from the dynamical system
point of view. The supposition of a homogeneous Universe is artificial,
and still presents a first generalization for its very primordial stages. 

Anyway, there is a well known conjecture that dissipative processes can take place in
a infinite dynamical system reducing the number of degrees of
freedom to just a few. In some sense, this argument reduces the artificiality of the
supposition of a homogeneous Universe.

Only the vacuum energy momentum classical source is considered in the full theory. It should be valid 
soon after the Planck era in which vacuum classical source seems the most natural condition.
The chaotic nature of the singularity is not addressed. 

There is not a unique way to totally characterize a singularity. In
this numerical investigation, the singularity is defined when 
$R_{ab}R^{ab}>10^6$.

It is found that for zero cosmological constant $\Lambda=0$, there are
basins that are   
attracted to zero, constant $4$-curvature solutions. This proves that
according to this theory, 
there are regions in the space of initial conditions, for which the
Minkowski geometry is structurally stable. That is, for considerably
anisotropic initial conditions isotropisation occurs,
and Minkowski space is obtained asymptotically, without the need for
fine tuning. The fact that the Minkowski space is an attractor to 
general Bianchi I spaces must be connected to non linear effects, 
see Appendix \ref{ap1}. 

On the other hand for a nonzero cosmological constant $\Lambda > 0$
the basins attract to a specific de Sitter solution. That is, constant
homogeneous, non null $4$-curvature solutions. Thus reproducing
Starobinsky's idea. 

Also, the complement of the set of initial conditions for the stable,
physically accepted solutions, for both cases $\Lambda=0$ and $\Lambda
>0$, form basins of attraction to 
singular, in some sense, Universes. Thus physically acceptable initial
conditions evolve in very few Planck times, to Universes with very high curvature
scalars. In this sense, this theory certainly is not a complete
one.  

It should be mentioned that the isotropisation process for non zero
cosmological constant is not a peculiarity of these higher derivatives
theories. It occurs in ordinary Einstein General Relativity theory. 
This is proved in the very interesting article \cite{wald}, where 
for $\Lambda >0$ all Bianchi models, except for highly positively curved
Bianchi IX become asymptotically de Sitter. The remarkable difference
between General Relativity theory and this higher derivatives theories
is that the isotropisation depends on the initial conditions. Isotropisation
is strongly dependent on initial conditions for quadratic curvature
theories such as \eqref{acao}.

The analytical solutions found by Barrow-Hervik and
Clifton-Barrow \cite{berkin} can be understood as limit sets in
the space of solutions of the quadratic theory \eqref{acao}. 
In particular, Barrow-Hervik found solutions for which isotropisation
does not occur. In this present work it is shown that the initial
conditions for which isotropistation does not occur, form a set with
non zero measure. Every {\it gray} point in FIG. \ref{stm} and
FIG. \ref{std} corresponds to solutions for which isotropisation does
not occur.A bifurcation analysis with respect to the parameters space, 
$\alpha,\;\beta,\;\Lambda$ is beyond the scope of the present work. 

\begin{acknowledgements}
S. D. P. Vitenti wishes to thank the Brazilian agency CAPES for
financial support. D. M. wishes to thank the Brazilian project {\it
Nova F\'\i sica no Espa\c co}.
\end{acknowledgements}
\appendix
\section{The differential equations \label{ap1}}
The following line element is chosen 
\[
ds^2=-dt^2+e^{2w_{{1}}(t)}dx^2+e^{2w_{{2}}(t)}dy^2+e^{2w_{{3}}(t)}dz^2.
\]
With this choice, the first time derivatives of the functions
$w_{{i}}(t)$ are related to the functions $a_{{i}}(t)$ of \eqref{elinha} in the
following manner
\begin{eqnarray}
y_{{1}}&=&\dot{w_{{1}}}=\frac{\dot{a_{{1}}}}{a_{{1}}}\\        
y_{{2}}&=&\dot{w_{{2}}}=\frac{\dot{a_{{2}}}}{a_{{2}}}\\       
y_{{3}}&=&\dot{w_{{3}}}=\frac{\dot{a_{{3}}}}{a_{{3}}}\\ 
y_{{4}}&=&\ddot{w_{{1}}}=\frac{\ddot{a_{{1}}}}{a_{{1}}}-\left(\frac{\dot{a_{{1}}}}{a_{{1}}} \right)^2\\        
y_{{5}}&=&\ddot{w_{{2}}}=\frac{\ddot{a_{{2}}}}{a_{{2}}}-\left(\frac{\dot{a_{{2}}}}{a_{{2}}} \right)^2\\       
y_{{6}}&=&\ddot{w_{{3}}}=\frac{\ddot{a_{{3}}}}{a_{{3}}}-\left(\frac{\dot{a_{{3}}}}{a_{{3}}} \right)^2\\                                 
y_{{7}}&=&\frac{d^3}{dt^3}w_{{1}}=\frac{1}{a_{{1}}}\frac{d^3}{dt^3}a_{{1}}-
\frac{\dot{a_{{1}}}}{a_{{1}}}\left(3\frac{\ddot{a_{{1}}}}{a_{{1}}}-2\left(\frac{\dot{a_{{1}}}}{a_{{1}}} \right)^2 \right)\\                                 
y_{{8}}&=&\frac{d^3}{dt^3}w_{{2}}=\frac{1}{a_{{2}}}\frac{d^3}{dt^3}a_{{2}}-
\frac{\dot{a_{{2}}}}{a_{{2}}}\left(3\frac{\ddot{a_{{2}}}}{a_{{2}}}-2\left(\frac{\dot{a_{{2}}}}{a_{{2}}} \right)^2 \right)\\                               
y_{{9}}&=&\frac{d^3}{dt^3}w_{{3}}=\frac{1}{a_{{3}}}\frac{d^3}{dt^3}a_{{3}}-
\frac{\dot{a_{{3}}}}{a_{{3}}}\left(3\frac{\ddot{a_{{3}}}}{a_{{3}}}-2\left(\frac{\dot{a_{{3}}}}{a_{{3}}}
\right)^2 \right).
\end{eqnarray}
The differential equations for the quadratic theory with the classical
source $T_{ab}=0$, \eqref{eqtotal}, are equivalent to 
\begin{eqnarray}
\frac{d^4}{dt^4}w_{{1}}&=&
{\frac {11}{9}}\,y_{{3}}y_{{1}}{y_{{2}}}^{2}-{\frac {19}{9}}\,y_{{3}}y
_{{2}}{y_{{1}}}^{2}+{\frac {11}{9}}\,y_{{2}}y_{{1}}{y_{{3}}}^{2}+5/3\,
y_{{1}}y_{{2}}y_{{5}}-13/3\,y_{{1}}y_{{2}}y_{{4}}\nonumber\\ 
&&-13/3\,y_{{1}}y_{{3}}
y_{{4}}+1/9\,y_{{1}}y_{{6}}y_{{2}}+5/3\,y_{{1}}y_{{3}}y_{{6}}+2/3\,y_{
{2}}y_{{3}}y_{{6}}-{\frac {14}{9}}\,y_{{2}}y_{{3}}y_{{4}}+2/3\,y_{{2}}
y_{{3}}y_{{5}}\nonumber\\ 
&&+1/9\,y_{{1}}y_{{3}}y_{{5}}- \left( {\frac {1}{108}}\,y_
{{3}}y_{{1}}{y_{{2}}}^{2}+{\frac {1}{108}}\,y_{{3}}y_{{2}}{y_{{1}}}^{2
}+{\frac {1}{108}}\,y_{{2}}y_{{1}}{y_{{3}}}^{2}+{\frac {1}{54}}\,y_{{1
}}y_{{6}}y_{{2}}\right.\nonumber\\
&&\left.+{\frac {1}{54}}\,y_{{2}}y_{{3}}y_{{4}}+{\frac {1}{54}
}\,y_{{1}}y_{{3}}y_{{5}}-{\frac {1}{108}}\,y_{{5}}y_{{4}}-{\frac {1}{
108}}\,y_{{6}}y_{{4}}-{\frac {1}{108}}\,y_{{6}}y_{{5}}+1/27\,{y_{{1}}}
^{2}{y_{{2}}}^{2}\right.\nonumber\\
&&\left.+1/27\,{y_{{1}}}^{2}{y_{{3}}}^{2}+1/27\,{y_{{2}}}^{2}
{y_{{3}}}^{2}-1/36\,{y_{{3}}}^{3}y_{{1}}-1/36\,{y_{{1}}}^{3}y_{{2}}-1/
36\,{y_{{1}}}^{3}y_{{3}}-1/36\,{y_{{2}}}^{3}y_{{1}}\right.\nonumber\\
&&\left.-1/36\,{y_{{3}}}^{3
}y_{{2}}-1/36\,{y_{{2}}}^{3}y_{{3}}-{\frac {1}{54}}\,y_{{1}}y_{{7}}-{
\frac {1}{54}}\,y_{{2}}y_{{8}}-{\frac {1}{54}}\,y_{{3}}y_{{9}}+{\frac 
{1}{108}}\,y_{{8}}y_{{1}}+{\frac
  {1}{108}}\,y_{{2}}y_{{7}}\right.\nonumber\\
&&\left.+{\frac {1}{
108}}\,y_{{9}}y_{{1}}+{\frac {1}{108}}\,y_{{9}}y_{{2}}+{\frac {1}{108}
}\,y_{{3}}y_{{8}}+{\frac {1}{108}}\,{y_{{4}}}^{2}+{\frac {1}{108}}\,{y
_{{5}}}^{2}+{\frac {1}{108}}\,{y_{{6}}}^{2}+{\frac {1}{108}}\,{y_{{2}}
}^{4}\right.\nonumber\\
&&\left.+{\frac {1}{108}}\,{y_{{3}}}^{4}-{\frac {1}{54}}\,{y_{{1}}}^{2}y_
{{4}}-{\frac {1}{54}}\,{y_{{2}}}^{2}y_{{5}}-{\frac {1}{54}}\,{y_{{3}}}
^{2}y_{{6}}+{\frac {1}{108}}\,y_{{3}}y_{{7}}+{\frac {1}{108}}\,{y_{{1}
}}^{4} \right) \alpha{\beta}^{-1}\nonumber\\
&&-\left( -1/18\,y_{{6}}-1/18\,y_{{5}}
+1/6\,\Lambda-1/18\,{y_{{3}}}^{2}-1/36\,y_{{3}}y_{{2}}-1/18\,{y_{{2}}
}^{2}-1/18\,y_{{4}}-1/36\,y_{{3}}y_{{1}}\right.\nonumber\\
&&\left.-1/18\,{y_{{1}}}^{2}-1/36\,y_{
{2}}y_{{1}}\right)/(G\beta)-{\frac {11}{9}}\,y_{{5}}y_{{4}}-{\frac {11}{9}}
\,y_{{6}}y_{{4}}+1/9\,y_{{6}}y_{{5}}-{\frac {5}{18}}\,{y_{{1}}}^{2}{y_
{{2}}}^{2}-{\frac {5}{18}}\,{y_{{1}}}^{2}{y_{{3}}}^{2}\nonumber\\
&&+{\frac {7}{18}}
\,{y_{{2}}}^{2}{y_{{3}}}^{2}+2/3\,{y_{{3}}}^{3}y_{{1}}-2/3\,{y_{{1}}}^
{3}y_{{2}}-2/3\,{y_{{1}}}^{3}y_{{3}}+2/3\,{y_{{2}}}^{3}y_{{1}}-{\frac 
{22}{9}}\,y_{{1}}y_{{7}}-1/9\,y_{{2}}y_{{8}}\nonumber\\
&&-1/9\,y_{{3}}y_{{9}}-1/9\,
y_{{8}}y_{{1}}-{\frac {16}{9}}\,y_{{2}}y_{{7}}-1/9\,y_{{9}}y_{{1}}+2/9
\,y_{{9}}y_{{2}}+2/9\,y_{{3}}y_{{8}}-{\frac {29}{18}}\,{y_{{4}}}^{2}
\nonumber\\
&&-{
\frac {5}{18}}\,{y_{{5}}}^{2}-{\frac {5}{18}}\,{y_{{6}}}^{2}-{\frac {5
}{18}}\,{y_{{2}}}^{4}-{\frac {5}{18}}\,{y_{{3}}}^{4}-1/9\,{y_{{1}}}^{2
}y_{{4}}-{\frac {7}{9}}\,{y_{{2}}}^{2}y_{{5}}-{\frac {7}{9}}\,{y_{{3}}
}^{2}y_{{6}}-{\frac {16}{9}}\,y_{{3}}y_{{7}}\nonumber\\
&&-4/3\,y_{{5}}{y_{{1}}}^{2}
-4/3\,y_{{6}}{y_{{1}}}^{2}+1/3\,y_{{6}}{y_{{2}}}^{2}+1/3\,{y_{{3}}}^{2
}y_{{5}}- \left(  \left( 4/3\,y_{{3}}y_{{1}}{y_{{2}}}^{2}-8/3\,y_{{3}}
y_{{2}}{y_{{1}}}^{2}\right.\right.\nonumber\\
&&\left.\left.+4/3\,y_{{2}}y_{{1}}{y_{{3}}}^{2}-4\,y_{{1}}y_{{2}
}y_{{5}}-4\,y_{{1}}y_{{2}}y_{{4}}-4\,y_{{1}}y_{{3}}y_{{4}}-4/3\,y_{{1}
}y_{{6}}y_{{2}}-4\,y_{{1}}y_{{3}}y_{{6}}+8\,y_{{2}}y_{{3}}y_{{6}}
\right.\right.\nonumber\\
&&\left.\left.+8/3
\,y_{{2}}y_{{3}}y_{{4}}+8\,y_{{2}}y_{{3}}y_{{5}}-4/3\,y_{{1}}y_{{3}}y_
{{5}}-4/3\,y_{{5}}y_{{4}}-4/3\,y_{{6}}y_{{4}}+8/3\,y_{{6}}y_{{5}}-8/3
\,{y_{{1}}}^{2}{y_{{2}}}^{2}\right.\right.\nonumber\\
&&\left.\left.-8/3\,{y_{{1}}}^{2}{y_{{3}}}^{2}+16/3\,{y_
{{2}}}^{2}{y_{{3}}}^{2}-4\,{y_{{1}}}^{3}y_{{2}}-4\,{y_{{1}}}^{3}y_{{3}
}+4\,{y_{{3}}}^{3}y_{{2}}+4\,{y_{{2}}}^{3}y_{{3}}-8/3\,y_{{1}}y_{{7}}
\right.\right.\nonumber\\
&&\left.\left.+
4/3\,y_{{2}}y_{{8}}+4/3\,y_{{3}}y_{{9}}-8/3\,y_{{8}}y_{{1}}+4/3\,y_{{2
}}y_{{7}}-8/3\,y_{{9}}y_{{1}}+4/3\,y_{{9}}y_{{2}}+4/3\,y_{{3}}y_{{8}}-
8/3\,{y_{{4}}}^{2}\right.\right.\nonumber\\
&&\left.\left.+4/3\,{y_{{5}}}^{2}+4/3\,{y_{{6}}}^{2}+4/3\,{y_{{2}}
}^{4}+4/3\,{y_{{3}}}^{4}-{\frac {32}{3}}\,{y_{{1}}}^{2}y_{{4}}+16/3\,{
y_{{2}}}^{2}y_{{5}}+16/3\,{y_{{3}}}^{2}y_{{6}}\right.\right.\nonumber\\
&&\left.\left.+4/3\,y_{{3}}y_{{7}}-4\,
y_{{5}}{y_{{1}}}^{2}-4\,y_{{6}}{y_{{1}}}^{2}+4\,y_{{6}}{y_{{2}}}^{2}+4
\,{y_{{3}}}^{2}y_{{5}}-8/3\,{y_{{1}}}^{4} \right) \beta+\left(-1/3\,
{y_{{3}}}^{2}-1/3\,y_{{5}}\right.\right.\nonumber\\
&&\left.\left.-1/3\,{y_{{2}}}^{2}-2/3\,y_{{3}}y_{{2}}-1/3
\,y_{{6}}+2/3\,{y_{{1}}}^{2}+2/3\,y_{{4}}+1/3\,y_{{2}}y_{{1}}+1/3\,y_{
{3}}y_{{1}}\right)/G \right) {\alpha}^{-1}\nonumber\\ 
&&+{\frac {7}{18}}\,{y_{{1}}}^{4}
\\
\frac{d^4}{dt^4}w_{{2}}
&=&-{\frac {19}{9}}\,y_{{3}}y_{{1}}{y_{{2}}}^{2}+{\frac {11}{9}}\,y_{{3}}
y_{{2}}{y_{{1}}}^{2}+{\frac {11}{9}}\,y_{{2}}y_{{1}}{y_{{3}}}^{2}-13/3
\,y_{{1}}y_{{2}}y_{{5}}+5/3\,y_{{1}}y_{{2}}y_{{4}}+2/3\,y_{{1}}y_{{3}}
y_{{4}}\nonumber\\
&&+1/9\,y_{{1}}y_{{6}}y_{{2}}+2/3\,y_{{1}}y_{{3}}y_{{6}}+5/3\,y_{
{2}}y_{{3}}y_{{6}}+1/9\,y_{{2}}y_{{3}}y_{{4}}-13/3\,y_{{2}}y_{{3}}y_{{
5}}-{\frac {14}{9}}\,y_{{1}}y_{{3}}y_{{5}}\nonumber
\end{eqnarray}
\newpage
\begin{eqnarray}
&&-\left( {\frac {1}{108}}\,y
_{{3}}y_{{1}}{y_{{2}}}^{2}+{\frac {1}{108}}\,y_{{3}}y_{{2}}{y_{{1}}}^{
2}+{\frac {1}{108}}\,y_{{2}}y_{{1}}{y_{{3}}}^{2}+{\frac {1}{54}}\,y_{{
1}}y_{{6}}y_{{2}}+{\frac {1}{54}}\,y_{{2}}y_{{3}}y_{{4}}\right.\nonumber\\
&&\left.+{\frac {1}{54
}}\,y_{{1}}y_{{3}}y_{{5}}-{\frac {1}{108}}\,y_{{5}}y_{{4}}-{\frac {1}{
108}}\,y_{{6}}y_{{4}}-{\frac
  {1}{108}}\,y_{{6}}y_{{5}}+1/27\,{y_{{1}}}
^{2}{y_{{2}}}^{2}+1/27\,{y_{{1}}}^{2}{y_{{3}}}^{2}\right.\nonumber\\
&&\left.+1/27\,{y_{{2}}}^{2}
{y_{{3}}}^{2}-1/36\,{y_{{3}}}^{3}y_{{1}}-1/36\,{y_{{1}}}^{3}y_{{2}}-1/
36\,{y_{{1}}}^{3}y_{{3}}-1/36\,{y_{{2}}}^{3}y_{{1}}-1/36\,{y_{{3}}}^{3
}y_{{2}}\right.\nonumber\\
&&\left.-1/36\,{y_{{2}}}^{3}y_{{3}}-{\frac {1}{54}}\,y_{{1}}y_{{7}}-{
\frac {1}{54}}\,y_{{2}}y_{{8}}-{\frac {1}{54}}\,y_{{3}}y_{{9}}+{\frac 
{1}{108}}\,y_{{8}}y_{{1}}+{\frac {1}{108}}\,y_{{2}}y_{{7}}+{\frac {1}{
108}}\,y_{{9}}y_{{1}}\right.\nonumber\\
&&\left.+{\frac
  {1}{108}}\,y_{{9}}y_{{2}}+{\frac {1}{108}
}\,y_{{3}}y_{{8}}+{\frac {1}{108}}\,{y_{{4}}}^{2}+{\frac {1}{108}}\,{y
_{{5}}}^{2}+{\frac {1}{108}}\,{y_{{6}}}^{2}+{\frac {1}{108}}\,{y_{{2}}
}^{4}+{\frac {1}{108}}\,{y_{{3}}}^{4}\right.\nonumber\\
&&\left.-{\frac {1}{54}}\,{y_{{1}}}^{2}y_
{{4}}-{\frac {1}{54}}\,{y_{{2}}}^{2}y_{{5}}-{\frac {1}{54}}\,{y_{{3}}}
^{2}y_{{6}}+{\frac {1}{108}}\,y_{{3}}y_{{7}}+{\frac {1}{108}}\,{y_{{1}
}}^{4} \right) \alpha{\beta}^{-1}-\left(-1/18\,y_{{6}}\right.\nonumber\\
&&\left.-1/18\,y_{{5}}
+1/6\,\Lambda-1/18\,{y_{{3}}}^{2}-1/36\,y_{{3}}y_{{2}}-1/18\,{y_{{2}}
}^{2}-1/18\,y_{{4}}-1/36\,y_{{3}}y_{{1}}\right.\nonumber\\
&&\left.-1/18\,{y_{{1}}}^{2}-1/36\,y_{
{2}}y_{{1}}\right)/(G\beta)-{\frac {11}{9}}\,y_{{5}}y_{{4}}+1/9\,y_{{6}}y_{{
4}}-{\frac {11}{9}}\,y_{{6}}y_{{5}}-{\frac {5}{18}}\,{y_{{1}}}^{2}{y_{
{2}}}^{2}\nonumber\\
&&+{\frac {7}{18}}\,{y_{{1}}}^{2}{y_{{3}}}^{2}-{\frac {5}{18}}
\,{y_{{2}}}^{2}{y_{{3}}}^{2}+2/3\,{y_{{1}}}^{3}y_{{2}}-2/3\,{y_{{2}}}^
{3}y_{{1}}+2/3\,{y_{{3}}}^{3}y_{{2}}-2/3\,{y_{{2}}}^{3}y_{{3}}\nonumber\\
&&-1/9\,y_
{{1}}y_{{7}}-{\frac {22}{9}}\,y_{{2}}y_{{8}}-1/9\,y_{{3}}y_{{9}}-{
\frac {16}{9}}\,y_{{8}}y_{{1}}-1/9\,y_{{2}}y_{{7}}+2/9\,y_{{9}}y_{{1}}
-1/9\,y_{{9}}y_{{2}}\nonumber\\
&&-{\frac {16}{9}}\,y_{{3}}y_{{8}}-{\frac {5}{18}}\,
{y_{{4}}}^{2}-{\frac {29}{18}}\,{y_{{5}}}^{2}-{\frac {5}{18}}\,{y_{{6}
}}^{2}+{\frac {7}{18}}\,{y_{{2}}}^{4}-{\frac {5}{18}}\,{y_{{3}}}^{4}-{
\frac {7}{9}}\,{y_{{1}}}^{2}y_{{4}}-1/9\,{y_{{2}}}^{2}y_{{5}}\nonumber\\
&&-{\frac {
7}{9}}\,{y_{{3}}}^{2}y_{{6}}+2/9\,y_{{3}}y_{{7}}+1/3\,y_{{6}}{y_{{1}}}
^{2}-4/3\,y_{{6}}{y_{{2}}}^{2}-{\frac {5}{18}}\,{y_{{1}}}^{4}-4/3\,{y_
{{2}}}^{2}y_{{4}}+1/3\,{y_{{3}}}^{2}y_{{4}}\nonumber\\
&&- \left(  \left( -8/3\,y_{{
3}}y_{{1}}{y_{{2}}}^{2}+4/3\,y_{{3}}y_{{2}}{y_{{1}}}^{2}+4/3\,y_{{2}}y
_{{1}}{y_{{3}}}^{2}-4\,y_{{1}}y_{{2}}y_{{5}}-4\,y_{{1}}y_{{2}}y_{{4}}+
8\,y_{{1}}y_{{3}}y_{{4}}\right.\right.\nonumber\\
&&\left.\left.-4/3\,y_{{1}}y_{{6}}y_{{2}}+8\,y_{{1}}y_{{3}}y
_{{6}}-4\,y_{{2}}y_{{3}}y_{{6}}-4/3\,y_{{2}}y_{{3}}y_{{4}}-4\,y_{{2}}y
_{{3}}y_{{5}}+8/3\,y_{{1}}y_{{3}}y_{{5}}-4/3\,y_{{5}}y_{{4}}\right.\right.\nonumber\\
&&\left.\left.+8/3\,y_{{
6}}y_{{4}}-4/3\,y_{{6}}y_{{5}}-8/3\,{y_{{1}}}^{2}{y_{{2}}}^{2}+16/3\,{
y_{{1}}}^{2}{y_{{3}}}^{2}-8/3\,{y_{{2}}}^{2}{y_{{3}}}^{2}+4\,{y_{{3}}}
^{3}y_{{1}}\right.\right.\nonumber\\
&&\left.\left.+4\,{y_{{1}}}^{3}y_{{3}}-4\,{y_{{2}}}^{3}y_{{1}}-4\,{y_{{2}
}}^{3}y_{{3}}+4/3\,y_{{1}}y_{{7}}-8/3\,y_{{2}}y_{{8}}+4/3\,y_{{3}}y_{{
9}}+4/3\,y_{{8}}y_{{1}}-8/3\,y_{{2}}y_{{7}}\right.\right.\nonumber\\
&&\left.\left.+4/3\,y_{{9}}y_{{1}}-8/3\,y
_{{9}}y_{{2}}+4/3\,y_{{3}}y_{{8}}+4/3\,{y_{{4}}}^{2}-8/3\,{y_{{5}}}^{2
}+4/3\,{y_{{6}}}^{2}-8/3\,{y_{{2}}}^{4}+4/3\,{y_{{3}}}^{4}\right.\right.\nonumber\\
&&\left.\left.+16/3\,{y_{{
1}}}^{2}y_{{4}}-{\frac {32}{3}}\,{y_{{2}}}^{2}y_{{5}}+16/3\,{y_{{3}}}^
{2}y_{{6}}+4/3\,y_{{3}}y_{{7}}+4\,y_{{6}}{y_{{1}}}^{2}-4\,y_{{6}}{y_{{
2}}}^{2}+4/3\,{y_{{1}}}^{4}\right.\right.\nonumber\\
&&\left.\left.-4\,{y_{{2}}}^{2}y_{{4}}+4\,{y_{{3}}}^{2}y_
{{4}} \right) \beta+\left(-1/3\,{y_{{3}}}^{2}+2/3\,y_{{5}}+2/3\,{y_{
{2}}}^{2}+1/3\,y_{{3}}y_{{2}}-1/3\,y_{{6}}\right.\right.\nonumber\\
&&\left.\left.-1/3\,{y_{{1}}}^{2}-1/3\,y_{
{4}}+1/3\,y_{{2}}y_{{1}}-2/3\,y_{{3}}y_{{1}}\right)/G \right) {\alpha}^{-1
}\\
\frac{d^4}{dt^4}w_{{3}}&=&
{\frac {11}{9}}\,y_{{3}}y_{{1}}{y_{{2}}}^{2}+{\frac {11}{9}}\,y_{{3}}y
_{{2}}{y_{{1}}}^{2}-{\frac {19}{9}}\,y_{{2}}y_{{1}}{y_{{3}}}^{2}+2/3\,
y_{{1}}y_{{2}}y_{{5}}+2/3\,y_{{1}}y_{{2}}y_{{4}}\nonumber\\
&&+5/3\,y_{{1}}y_{{3}}y_
{{4}}-{\frac {14}{9}}\,y_{{1}}y_{{6}}y_{{2}}-13/3\,y_{{1}}y_{{3}}y_{{6
}}-13/3\,y_{{2}}y_{{3}}y_{{6}}+1/9\,y_{{2}}y_{{3}}y_{{4}}+5/3\,y_{{2}}
y_{{3}}y_{{5}}\nonumber\\
&&+1/9\,y_{{1}}y_{{3}}y_{{5}}- \left( {\frac {1}{108}}\,y_
{{3}}y_{{1}}{y_{{2}}}^{2}+{\frac {1}{108}}\,y_{{3}}y_{{2}}{y_{{1}}}^{2
}+{\frac {1}{108}}\,y_{{2}}y_{{1}}{y_{{3}}}^{2}+{\frac {1}{54}}\,y_{{1
}}y_{{6}}y_{{2}}\right.\nonumber
\end{eqnarray}
\begin{eqnarray}
&&\left.+{\frac {1}{54}}\,y_{{2}}y_{{3}}y_{{4}}+{\frac {1}{54}
}\,y_{{1}}y_{{3}}y_{{5}}-{\frac {1}{108}}\,y_{{5}}y_{{4}}-{\frac {1}{
108}}\,y_{{6}}y_{{4}}-{\frac {1}{108}}\,y_{{6}}y_{{5}}+1/27\,{y_{{1}}}
^{2}{y_{{2}}}^{2}\right.\nonumber\\
&&\left.+1/27\,{y_{{1}}}^{2}{y_{{3}}}^{2}+1/27\,{y_{{2}}}^{2}
{y_{{3}}}^{2}-1/36\,{y_{{3}}}^{3}y_{{1}}-1/36\,{y_{{1}}}^{3}y_{{2}}-1/
36\,{y_{{1}}}^{3}y_{{3}}-1/36\,{y_{{2}}}^{3}y_{{1}}\right.\nonumber\\
&&\left.-1/36\,{y_{{3}}}^{3
}y_{{2}}-1/36\,{y_{{2}}}^{3}y_{{3}}-{\frac {1}{54}}\,y_{{1}}y_{{7}}-{
\frac {1}{54}}\,y_{{2}}y_{{8}}-{\frac {1}{54}}\,y_{{3}}y_{{9}}+{\frac 
{1}{108}}\,y_{{8}}y_{{1}}+{\frac {1}{108}}\,y_{{2}}y_{{7}}\right.\nonumber\\
&&\left.+{\frac {1}{
108}}\,y_{{9}}y_{{1}}+{\frac {1}{108}}\,y_{{9}}y_{{2}}+{\frac {1}{108}
}\,y_{{3}}y_{{8}}+{\frac {1}{108}}\,{y_{{4}}}^{2}+{\frac {1}{108}}\,{y
_{{5}}}^{2}+{\frac {1}{108}}\,{y_{{6}}}^{2}+{\frac {1}{108}}\,{y_{{2}}
}^{4}\right.\nonumber\\
&&\left.+{\frac {1}{108}}\,{y_{{3}}}^{4}-{\frac {1}{54}}\,{y_{{1}}}^{2}y_
{{4}}-{\frac {1}{54}}\,{y_{{2}}}^{2}y_{{5}}-{\frac {1}{54}}\,{y_{{3}}}
^{2}y_{{6}}+{\frac {1}{108}}\,y_{{3}}y_{{7}}+{\frac {1}{108}}\,{y_{{1}
}}^{4} \right) \alpha{\beta}^{-1}-\left(-1/18\,y_{{6}}\right.\nonumber\\
&&\left.-1/18\,y_{{5}}
+1/6\,\Lambda-1/18\,{y_{{3}}}^{2}-1/36\,y_{{3}}y_{{2}}-1/18\,{y_{{2}}
}^{2}-1/18\,y_{{4}}-1/36\,y_{{3}}y_{{1}}\right.\nonumber\\
&&\left.-1/18\,{y_{{1}}}^{2}-1/36\,y_{
{2}}y_{{1}}\right)/(G\beta)+1/9\,y_{{5}}y_{{4}}-{\frac {11}{9}}\,y_{{6}}y_{{
4}}-{\frac {11}{9}}\,y_{{6}}y_{{5}}+{\frac {7}{18}}\,{y_{{1}}}^{2}{y_{
{2}}}^{2}-{\frac {5}{18}}\,{y_{{1}}}^{2}{y_{{3}}}^{2}\nonumber\\
&&-{\frac {5}{18}}
\,{y_{{2}}}^{2}{y_{{3}}}^{2}-2/3\,{y_{{3}}}^{3}y_{{1}}+2/3\,{y_{{1}}}^
{3}y_{{3}}-2/3\,{y_{{3}}}^{3}y_{{2}}+2/3\,{y_{{2}}}^{3}y_{{3}}-1/9\,y_
{{1}}y_{{7}}-1/9\,y_{{2}}y_{{8}}\nonumber\\
&&-{\frac {22}{9}}\,y_{{3}}y_{{9}}+2/9\,
y_{{8}}y_{{1}}+2/9\,y_{{2}}y_{{7}}-{\frac {16}{9}}\,y_{{9}}y_{{1}}-{
\frac {16}{9}}\,y_{{9}}y_{{2}}-1/9\,y_{{3}}y_{{8}}-{\frac {5}{18}}\,{y
_{{4}}}^{2}-{\frac {5}{18}}\,{y_{{5}}}^{2}\nonumber\\
&&-{\frac {29}{18}}\,{y_{{6}}}
^{2}-{\frac {5}{18}}\,{y_{{2}}}^{4}+{\frac {7}{18}}\,{y_{{3}}}^{4}-{
\frac {7}{9}}\,{y_{{1}}}^{2}y_{{4}}-{\frac {7}{9}}\,{y_{{2}}}^{2}y_{{5
}}-1/9\,{y_{{3}}}^{2}y_{{6}}-1/9\,y_{{3}}y_{{7}}+1/3\,y_{{5}}{y_{{1}}}
^{2}\nonumber\\
&&-4/3\,{y_{{3}}}^{2}y_{{5}}- \left(  \left( 4/3\,y_{{3}}y_{{1}}{y_{
{2}}}^{2}+4/3\,y_{{3}}y_{{2}}{y_{{1}}}^{2}-8/3\,y_{{2}}y_{{1}}{y_{{3}}
}^{2}+8\,y_{{1}}y_{{2}}y_{{5}}+8\,y_{{1}}y_{{2}}y_{{4}}
\right.\right.\nonumber\\
&&\left.\left.-4\,y_{{1}}y_{{
3}}y_{{4}}+8/3\,y_{{1}}y_{{6}}y_{{2}}-4\,y_{{1}}y_{{3}}y_{{6}}-4\,y_{{
2}}y_{{3}}y_{{6}}-4/3\,y_{{2}}y_{{3}}y_{{4}}-4\,y_{{2}}y_{{3}}y_{{5}}-
4/3\,y_{{1}}y_{{3}}y_{{5}}\right.\right.\nonumber\\
&&\left.\left.+8/3\,y_{{5}}y_{{4}}-4/3\,y_{{6}}y_{{4}}-4/3
\,y_{{6}}y_{{5}}+16/3\,{y_{{1}}}^{2}{y_{{2}}}^{2}-8/3\,{y_{{1}}}^{2}{y
_{{3}}}^{2}-8/3\,{y_{{2}}}^{2}{y_{{3}}}^{2}\right.\right.\nonumber\\
&&\left.\left.-4\,{y_{{3}}}^{3}y_{{1}}+4
\,{y_{{1}}}^{3}y_{{2}}+4\,{y_{{2}}}^{3}y_{{1}}-4\,{y_{{3}}}^{3}y_{{2}}
+4/3\,y_{{1}}y_{{7}}+4/3\,y_{{2}}y_{{8}}-8/3\,y_{{3}}y_{{9}}+4/3\,y_{{
8}}y_{{1}}\right.\right.\nonumber\\
&&\left.\left.+4/3\,y_{{2}}y_{{7}}+4/3\,y_{{9}}y_{{1}}+4/3\,y_{{9}}y_{{2}}
-8/3\,y_{{3}}y_{{8}}+4/3\,{y_{{4}}}^{2}+4/3\,{y_{{5}}}^{2}-8/3\,{y_{{6
}}}^{2}+4/3\,{y_{{2}}}^{4}\right.\right.\nonumber\\
&&\left.\left.-8/3\,{y_{{3}}}^{4}+16/3\,{y_{{1}}}^{2}y_{{4
}}+16/3\,{y_{{2}}}^{2}y_{{5}}-{\frac {32}{3}}\,{y_{{3}}}^{2}y_{{6}}-8/
3\,y_{{3}}y_{{7}}+4\,y_{{5}}{y_{{1}}}^{2}-4\,{y_{{3}}}^{2}y_{{5}}
\right.\right.\nonumber\\
&&\left.\left.+4/3
\,{y_{{1}}}^{4}+4\,{y_{{2}}}^{2}y_{{4}}-4\,{y_{{3}}}^{2}y_{{4}}
 \right) \beta+\left(2/3\,{y_{{3}}}^{2}-1/3\,y_{{5}}-1/3\,{y_{{2}}}^
{2}+1/3\,y_{{3}}y_{{2}}+2/3\,y_{{6}}\right.\right.\nonumber\\
&&\left.\left.-1/3\,{y_{{1}}}^{2}-1/3\,y_{{4}}-2
/3\,y_{{2}}y_{{1}}+1/3\,y_{{3}}y_{{1}}\right)/G \right) {\alpha}^{-1}-{
\frac{5}{18}}\,{y_{{1}}}^{4}+1/3\,{y_{{2}}}^{2}y_{{4}}\nonumber\\
&&-4/3\,{y_{{3}}}^{2}y_{{4}}
\end{eqnarray}

\end{document}